\documentclass[preprint,12pt]{elsarticle}

\usepackage{amssymb}
\usepackage{comment}
\usepackage{amsmath}
\newcommand\Rey{\mbox{\textit{Re}}}  
\newcommand{\vect}[1]{\boldsymbol{#1}}
\newcommand\Fro{\mbox{\textit{Fr}}}  

\journal{Computers \& Fluids}

\begin{document}

\begin{frontmatter}



\title{Leveraging modal structure similarity for simulation of spatially evolving wakes}


\author[1]{Divyanshu Gola}
\author[2]{Sutanu Sarkar}

\affiliation[1]{organization={Center for Naval Research \& Education},
            addressline={University of Michigan}, 
            city={Ann Arbor},
            postcode={48105}, 
            state={Michigan},
            country={USA}}

\affiliation[2]{organization={Department of Mechanical and Aerospace Engineering},
            addressline={University of California San Diego}, 
            city={La Jolla},
            postcode={92092}, 
            state={California},
            country={USA}}



\begin{abstract}
We present a new methodology to enable efficient simulation of high Reynolds number wakes. In this approach, a body-exclusive hybrid simulation at $\Rey = 5 \times 10^4$ is initialized using inflow fields reconstructed from a lower Reynolds number ($\Rey = 5 \times 10^3$) body-inclusive simulation. Spectral Proper Orthogonal Decomposition (SPOD) is employed to identify dominant coherent structures, and a low-rank reconstruction generates physically meaningful inflow conditions. The resulting  simulations accurately recover key large-scale flow properties, including the vortex shedding mode, and match the entire spectral content of a fully body-inclusive $\Rey = 5 \times 10^4$ reference simulation beyond an adjustment region. SPOD eigenspectra confirm agreement across both low and high frequencies. This strategy achieves over an order-of-magnitude reduction in computational cost by avoiding direct high-Re body-inclusive simulations, offering a scalable framework for simulating complex wakes using low $\Rey$ as well as reduced-order inflow prescriptions.
\end{abstract}

\begin{graphicalabstract}
\includegraphics[width=\linewidth, keepaspectratio]{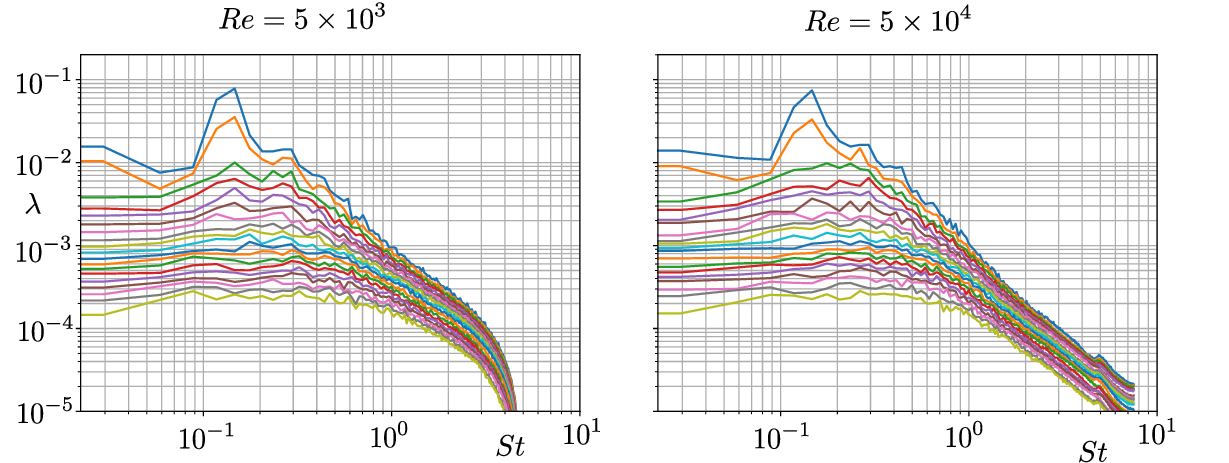}
\end{graphicalabstract}

\begin{highlights}
\item Low frequency coherent structures of a turbulent wake show Reynolds number independence
\item This coherent structure independence can be used to make high Reynolds number simulations computationally cheaper
\end{highlights}

\begin{keyword}

Modal decomposition \sep hybrid simulations \sep turbulent wakes



\end{keyword}

\end{frontmatter}



\section{Introduction}

The advancement of computational capabilities has greatly expanded the frontiers of fluid dynamics research, but simulating high Reynolds number ($Re$) flows especially those involving wake evolution over long distances remains computationally expensive. Direct Numerical Simulation (DNS) and Large Eddy Simulation (LES) at high $Re$ are often constrained by the required resolution and domain size, leading to impractical runtimes and resource demands. This challenge is particularly pronounced in studies of bluff body wakes, where the far-wake region exhibits behavior that requires large computational domains to capture the essential physics. As interest grows in the high-$\Rey$ late-wake dynamics of bluff bodies, the demand for efficient simulation strategies becomes more urgent.

Hybrid simulation methods have emerged as a powerful tool to alleviate the computational cost associated with large-domain, high-$Re$ wake simulations. In the present context, a hybrid simulation strategy decomposes the overall problem into two subdomains: a high-resolution, body-inclusive simulation with a limited downstream extent; and a lower-resolution, body-exclusive simulation with an extended computational domain. Flow data from the near wake of the former are used to prescribe inflow boundary conditions for the latter. This approach significantly reduces the computational expense by allowing a coarser mesh and larger timestep in the downstream domain, while still capturing essential flow features initiated by the presence of the body. Previous studies have successfully employed hybrid simulations to investigate unsteady wake dynamics of spheres and spheroids~\cite{Pasquetti:2011,vandine_hybrid_2018,ortiz2021high}. However, at sufficiently high Reynolds numbers, even the short-domain body-inclusive simulation can become prohibitively expensive. Enhancing the cost-effectiveness of hybrid simulations especially at high $Re$ is therefore the primary objective of this work.

A key component of this effort is the use of Spectral Proper Orthogonal Decomposition (SPOD)~\cite{towne_spectral_2018}, a technique that identifies coherent structures from spatio-temporal data by solving a frequency-resolved eigenvalue problem. SPOD provides a natural framework to analyze and rank flow features by energy content at each frequency. For stratified wakes, SPOD has proven particularly insightful: Nidhan et al.~\cite{nidhan_analysis_2022} applied SPOD to identify dominant vortex-shedding modes in the wake of a disk at $Re = 5 \times 10^4$, demonstrating the role of coherent structures in wake persistence and transition. More recently, SPOD-based approaches have been extended to reconstruct incomplete or gappy data~\cite{nekkanti2023gappy}, opening avenues for model reduction and efficient inflow generation.

This study explores the following central question: Can a low-$Re$ body-inclusive simulation be used to construct accurate inflow boundary conditions for a high-$Re$ body-exclusive simulation? If so, what are the minimal flow features required in the low-$Re$ simulation to achieve this fidelity? We investigate these questions using unstratified disk wakes at $Re = 5 \times 10^3$ and $Re = 5 \times 10^4$ as canonical test cases. The methodology is described in Section~\ref{methods}. In Section~\ref{res:5k_50k_bi}, we employ SPOD to assess similarities in the coherent structures between the low and high $Re$ body-inclusive simulations. Section~\ref{res:50kbe} presents results from the high $Re$ body-exclusive simulations initialized using inflow data from the low $Re$ run, including low-rank reconstructions. These are compared against the full-domain $Re = 5 \times 10^4$ reference simulation by Chongsiripinyo and Sarkar~\cite{chongsiripinyo_decay_2020}. Finally, we discuss the implications and limitations of this SPOD-based hybrid approach in Section~\ref{res:discuss}.

\section{Methodology}
\label{methods}

\subsection{Hybrid simulations}\label{sec:hyb_sim_method}

\begin{figure}

\centerline{\includegraphics[width=\linewidth, keepaspectratio]{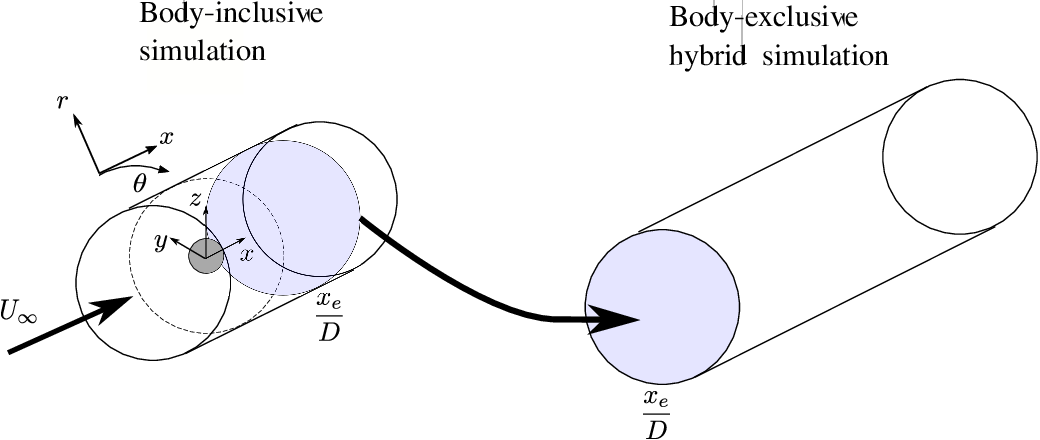}}
\caption{Hybrid simulation setup. Unlike the typical hybrid simulations, the body-inclusive simulation is at a lower $Re$ than the body-exclusive simulation.}
\label{fig:hybrid}

\end{figure}

The concept of a hybrid simulation involves decomposing a large simulation (in the spatial sense) into two smaller, more manageable sub-simulations. This approach is particularly advantageous in body-inclusive simulations, where computational costs are elevated due to the high resolution required in the vicinity of the body to resolve its thin boundary layer and the near wake. In a typical hybrid setup, the first simulation includes the body but is confined to a limited spatial extent usually in the streamwise direction. Once statistical stationarity is achieved, data from this body-inclusive simulation, often referred to as \emph{hybrid planes}, are extracted at a downstream location and used as inflow boundary conditions for the second, body-exclusive simulation. The absence of the body in the latter permits the use of coarser resolution and a significantly larger spatial domain. This, in turn, allows for a larger time step and faster attainment of statistical stationarity. Thus, a hybrid simulation effectively makes the problem less stiff by splitting it into two separate problems, and also allowing for less runtime memory usage. When properly configured, the two-part hybrid simulation can deliver computational efficiency and reduced costs without compromising the accuracy relative to a full-domain body-inclusive simulation.

Selecting an appropriate decomposition or ``break-up'' point is critical. In this study, we focus on streamwise decomposition by choosing a downstream location relative to the body to terminate the first simulation and extract hybrid planes for initializing the second. The extraction point must strike a balance: it should not be placed too far downstream, which would negate the computational benefits, nor too close to the wake, which may result in highly variable inflow velocity rendering inaccurate the assumption of spatially constant advection of the inflow plane data over a time step. VanDine et al.~\cite{vandine_hybrid_2018}, in their direct numerical simulation (DNS) of a sphere, identified \( x/D = 9 \) (nine body diameters downstream) as an optimal extraction location. Based on this finding, all hybrid simulations in the present work adopt an extraction location of \( x/D = 10 \), where \( D \) denotes the disk diameter.

A distinguishing feature of the hybrid simulations presented in this study is that the second, body-exclusive simulation is performed at a Reynolds number ten times higher than that of the body-inclusive simulation. The rationale behind this Reynolds number upscaling is provided \textit{a priori} in Sections~\ref{sec:spod_method} and~\ref{sec:lrr_method}, with \textit{a posteriori} validation presented in Section~\ref{res:50kbe}. A schematic of the hybrid simulation framework is shown in Figure~\ref{fig:hybrid}. Large Eddy Simulations (LES) are employed for both the body-inclusive and body-exclusive simulations. The disk, with diameter \( D \) and thickness \( 0.01D \), is positioned at \( x/D = 0 \) and \( r/D = 0 \), perpendicular to the freestream with velocity \( U_{\infty} \), in the body-inclusive simulation (referred to as 5kBI) at \( \Rey = 5 \times 10^3 \). The radial extent of the domain is set to \( L_r = 15D \) in both simulations. The streamwise domain for 5kBI ranges from \( L_{x-} = -30D \) to \( L_{x+} = 15D \), and hybrid planes are extracted at \( x/D = 10 \) at intervals of 0.013 non-dimensional time units. 

The body-exclusive simulations, referred to as SPOD-reconstructed hybrid (SRH) simulations, share the same radial and azimuthal extent and resolution as 5kBI, but their streamwise extent follows the configuration used by Chongsiripinyo et al.~\cite{chongsiripinyo_decay_2020}, ranging from \( L_{x-} = 10D \) to \( L_{x+} = 120D \). The various SRH simulation cases are described in detail in Section~\ref{sec:lrr_method}, and the simulation parameters are summarized in Table~\ref{tab:phys_parameters_2}, with details of the simulation model in Section \ref{model}.

\begin{table}

\begin{center}
\begin{tabular}{lccccccccc }

    Case  &    $Re$      & Type                    & $L_r$  & $L_{\theta}$ & $L_{x-}$ & $L_{x+}$ & $N_{r}$ & $N_{\theta}$ & $N_x$ \\
\hline
    5kBI  &    $5 \times 10^3$    & Body-inclusive     &  $15$  & $2\pi$   & $-30$           & $15$   & $364$ & $256$ & $1538$  \\
    SRH-N1 & $5 \times 10^4$   & Body-exclusive     &  $15$  & $2\pi$   & $10$  &         $120$   &  $364$ & $256$ & $3074$  \\
   SRH-N2 & $5 \times 10^4$   & Body-exclusive     &  $15$  & $2\pi$   & $10$  &         $120$   &  $364$ & $256$ & $3074$  \\ 
   SRH-N3 & $5 \times 10^4$   & Body-exclusive     &  $15$  & $2\pi$   & $10$  &         $120$   &  $364$ & $256$ & $3074$  \\
   SRH-N6 & $5 \times 10^4$   & Body-exclusive     &  $15$  & $2\pi$   & $10$  &         $120$   &  $364$ & $256$ & $3074$  \\
   SRH-St0103 & $5 \times 10^4$   & Body-exclusive     &  $15$  & $2\pi$   & $10$  &         $120$   &  $364$ & $256$ & $3074$  \\
   SRH-NAll & $5 \times 10^4$   & Body-exclusive     &  $15$  & $2\pi$   & $10$  &         $120$   &  $364$ & $256$ & $3074$  \\
   50kBI & $5 \times 10^4$   & Body-inclusive    &  $15$  & $2\pi$   & $-30$  &         $120$   &  $364$ & $256$ & $4608$ \\

& & (for comparison)

\end{tabular}

\end{center}

\caption{Physical parameters of all simulation cases (50kBI by Chongsiripinyo et al.\cite{chongsiripinyo_decay_2020})}
\label{tab:phys_parameters_2}
\end{table}

\subsection{Spectral Proper Orthogonal Decomposition}\label{sec:spod_method}

In fluid dynamics, Spectral Proper Orthogonal Decomposition (SPOD) has been widely employed to extract coherent structures from turbulent flows~\citep{nidhan_analysis_2022, towne_spectral_2018}. The core idea of SPOD is to derive basis functions for the spatio-temporal flow field that optimally capture the kinetic energy content. Mathematically, if $\vect{Q}(\vect{x},t) = [u_{i}'(\vect{x},t)]^{T}$ denotes the fluctuating velocity field over the domain $\Pi$, SPOD seeks deterministic functions $\vect{\Psi}(\vect{x},t)$ that maximize the ensemble-averaged projection onto $\vect{Q}(\vect{x},t)$, i.e.,

\begin{equation}
\max_{\vect{\Psi}} \frac{\langle | \{ \vect{Q}(\vect{x},t), \vect{\Psi}(\vect{x},t) \} |^{2} \rangle}{\{ \vect{\Psi}(\vect{x},t), \vect{\Psi}(\vect{x},t) \}},
\end{equation}

where $\langle \cdot \rangle$ denotes the ensemble average. For turbulent wakes, the inner product $\{ \vect{Q}^{(1)}(\vect{x},t), \vect{Q}^{(2)}(\vect{x},t) \}$ is defined as

\begin{equation}
\{ \vect{Q}^{(1)}(\vect{x},t), \vect{Q}^{(2)}(\vect{x},t) \} = \int_{-\infty}^{\infty} \int_{\Pi} \vect{Q}^{(2)*}(\vect{x},t)\, \mathbf{W} \, \vect{Q}^{(1)}(\vect{x},t) \, d\vect{x} \, dt,
\end{equation}

where the asterisk denotes the conjugate transpose of the original matrix, and $\mathbf{W}$ is a positive-definite Hermitian matrix, which for this study, has dimensions $N_r \times N_r$, containing the quadrature weights corresponding to the radial grid used in the simulation.

In this study, SPOD is utilized to extract the eigenspectra of the velocity field in the near to intermediate wake region downstream of the disk, with comparisons drawn across two different Reynolds numbers. The numerical implementation of SPOD is performed on the time-resolved simulation data at  two-dimensional cross-stream slices of the velocity field.   The mean-subtracted dataset, comprising $N$ snapshots, is divided into $N_{\text{blk}}$ blocks with an overlap of $N_{\text{ovlp}}$ snapshots per block. Each block contains $N_{\text{freq}}$ entries, and the data matrix is structured as:

\begin{equation}
\vect{Q} = [\vect{q}^{(1)}, \vect{q}^{(2)}, \vect{q}^{(3)}, \ldots, \vect{q}^{(N_{\text{freq}})}],
\end{equation}

where each snapshot vector is defined as $\vect{q}^{(i)} = [\vect{u}'^{(i)}, \rho'^{(i)}]^{T}$. Assuming statistical stationarity, a discrete Fourier transform is applied in time to each block, resulting in $N_{\text{blk}}$ Fourier realizations $\hat{\vect{Q}}_f$ for each frequency $f$. The SPOD eigenvalues and eigenvectors at a given frequency are then obtained by solving the eigenvalue problem:

\begin{equation}
\hat{\vect{Q}}_f^* \mathbf{W} \hat{\vect{Q}}_f \boldsymbol{\Gamma}_f = \boldsymbol{\Gamma}_f \boldsymbol{\Lambda}_f,
\end{equation}

where $\boldsymbol{\Lambda}_f = \text{diag}(\lambda_f^{(1)}, \lambda_f^{(2)}, \ldots, \lambda_f^{(N_{\text{blk}})})$ is a diagonal matrix containing the SPOD eigenvalues sorted in descending order of energy content. The SPOD modes at each frequency are then computed as:

\begin{equation}
\boldsymbol{\Phi}_f = \hat{\vect{Q}}_f \boldsymbol{\Gamma}_f \boldsymbol{\Lambda}_f^{-1/2}.
\end{equation}

This study focuses on the behavior of the SPOD spectra, defined by the magnitude of the $N_{\text{blk}}$ eigenvalues at each frequency $f$, as a function of the non-dimensional frequency, $St$. The variations in spectral content are analyzed across different flow cases with varying Froude numbers, $\Fro$. For a detailed discussion on SPOD and its application to general and stratified flows, readers are referred to Towne et al.~\citep{towne_spectral_2018} and Nidhan et al.~\citep{nidhan_analysis_2022}, respectively.

\subsection{Low-Rank Reconstruction}\label{sec:lrr_method}

Since SPOD modes are ranked by their energy content (eigenvalues), the leading modes capture the dominant energetic features of the flow field. This naturally raises the question of how many leading SPOD modes are required to reconstruct key features of the flow, such as Reynolds stresses or turbulent kinetic energy (TKE). Nidhan et al.~\citep{nidhan_spectral_2020} addressed this question for the wake behind a disk at $Re = 5 \times 10^4$, demonstrating that as few as three leading modes were sufficient to reconstruct at least 90\% of the Reynolds stress $\langle -u_{x}' u_{r}' \rangle$, although only about 50\% of the TKE was captured. Their study also included azimuthal decomposition to investigate the contributions of specific azimuthal modes (e.g., $m=1$ and $m=2$). However, the present work does not use azimuthal decomposition, as the proposed reconstruction method is also intended for flows lacking axisymmetry, such as those involving angle of attack or density stratification.

The reconstruction of inlet flow planes for the hybrid simulations in this study is based on a frequency-domain approach, which is the exact inverse of the SPOD formulation outlined in Section~\ref{sec:spod_method}, and described in detail by Nekkanti and Schmidt~\citep{nekkanti_frequency_2021}. This inverse process is implemented by manipulating the associated expansion coefficients. The original Fourier-transformed realizations at each frequency can be reconstructed via

\begin{equation}
    \hat{\mathbf{Q}}_f = \boldsymbol{\Phi}_f \mathbf{A}_f,
\end{equation}

where $\mathbf{A}_f$ is the matrix of expansion coefficients. These coefficients are computed as

\begin{equation}
    \mathbf{A}_f = \boldsymbol{\Lambda}_f^{1/2} \boldsymbol{\Psi}_f^* = \boldsymbol{\Phi}_f^* \mathbf{W} \hat{\mathbf{Q}}_f.
\end{equation}

As shown, the expansion coefficients can either be stored during SPOD computation or reconstructed later by projecting the Fourier realizations onto the SPOD modes.

Each column of the matrix
\begin{equation}
\mathbf{A}_f =
\begin{bmatrix}
a_{11} & a_{12} & \cdots & a_{1\,N_{\text{blk}}} \\
a_{21} & a_{22} & \cdots & a_{2\,N_{\text{blk}}} \\
\vdots & \vdots & \ddots & \vdots \\
a_{N_{\text{blk}}\,1} & a_{N_{\text{blk}}\,2} & \cdots & a_{N_{\text{blk}}\,N_{\text{blk}}}
\end{bmatrix}
\end{equation}
contains the expansion coefficients needed to reconstruct individual Fourier-domain realizations using the SPOD modes. Conversely, each row corresponds to an SPOD mode represented as a linear combination of these realizations. This reciprocal structure is further demonstrated by rewriting the reconstruction equation as

\begin{equation}
    \boldsymbol{\Phi}_f = \hat{\mathbf{Q}}_f \mathbf{A}_f^*.
\end{equation}

The number of retained SPOD modes is controlled by zeroing out the corresponding rows in $\mathbf{A}_f$ for the excluded modes. To mitigate edge effects caused by windowing, the first half of the first block and the last half of the last block are excluded from the reconstructed dataset, as recommended by Nekkanti and Schmidt~\citep{nekkanti_frequency_2021}.

In this study, SPOD and its inverse are applied to a total of 25,600 two-dimensional snapshots extracted from the $5$kBI simulation at $x/D = 10$. These snapshots are separated by a time increment of $U_{\infty} \Delta t / D = 0.013$. Due to memory constraints, the full dataset is divided into five groups of $N_{\text{blk}} = 5120$ snapshots, each comprising every fifth snapshot from the original set. This effectively results in a new time step of $U_{\infty} \Delta t / D = 0.065$. For the decomposition, the overlap is set to $N_{\text{ovlp}} = 256$ and the block length to $N_{\text{freq}} = 512$.

The importance of leading-order modes is underscored in Section~\ref{res:5k_50k_bi}, where it is shown that the dominant SPOD modes are consistent across both the $5$kBI and $50$kBI cases. This observation motivates the investigation of whether low-rank modes extracted at low Reynolds number can be used to initialize simulations at higher Reynolds numbers. To explore this, we evaluate reconstructions using 1, 2, 3, and 6 leading SPOD modes designated as cases SRH-N1, SRH-N2, SRH-N3, and SRH-N6, respectively.

Two additional cases are also studied. The first, SRH-NAll, uses all available SPOD modes, effectively replicating the $5$kBI field. The second, SRH-St0103, applies a band-pass filter to retain only modes within the vortex-shedding frequency range ($0.1 < St < 0.3$) around $St = 0.146$. This case aims to test whether a narrow spectral band (commonly obtained in experimental measurements) can effectively initialize higher-$Re$ simulations.

The body-inclusive simulation of a disk at $Re = 5 \times 10^4$ conducted by Chongsiripinyo et al.~\citep{chongsiripinyo_decay_2020} serves as the benchmark for validation and is referred to as 50kBI throughout this work. The simulation configurations are summarized in Table~\ref{tab:phys_parameters_2}.



\subsection{Simulation model}
\label{model}

The filtered nondimensional Navier Stokes equations are solved on a cylindrical grid:

\begin{equation} 
\frac{\partial u_{i}}{\partial x_{i}} = 0,
\label{conservation_eqn}
\end{equation}

\begin{equation} 
\frac{\partial u_{i}}{\partial t} + u_{j}\frac{\partial u_{i}}{\partial x_{j}} = -\frac{\partial p}{\partial x_{i}} + \frac{1}{Re} \frac{\partial}{\partial x_{j}}\Big[(1 + \frac{\nu_{sgs}}{\nu}) \frac{\partial u_{i}}{\partial x_{j}}\Big]  
\label{momentum_eqn}
\end{equation}

A dynamic Smagorinsky model (\citet{Germano1991}) is used to compute the subgrid-scale viscosity, given by \(\nu_{\text{sgs}} = C\tilde{\Delta}^2 |\tilde{S}|\), where the coefficient \(C\) is obtained using Lilly’s least-squares method \citep{Lilly1992} combined with the Lagrangian averaging approach of \citet{Meneveau1996}. Both the molecular and subgrid-scale Prandtl numbers are set to unity. Spatial derivatives are discretized using second-order central differences. Time integration employs a low-storage third-order Runge--Kutta scheme for the advective terms and a Crank--Nicolson scheme for viscous terms. The disk in 5kBI is modeled using the immersed boundary method described by \citet{Balaras2004} and \citet{Yang2006}. A uniform inflow is applied at the inlet ($x/D = L_{x-}$) for 5kBI, while an Orlanski-type convective boundary condition \citep{Orlanski1976} is imposed at the outlet ($x/D = L_{x+}$) for all cases. Neumann boundary conditions are used at the radial boundary (r/D = $L_{r+}$). At the centerline (\(r = 0\)), radial and azimuthal velocity components are defined using symmetric averaging of ghost cells. Statistics are obtained by time-averaging once the flow reaches a statistical steady state.

\section{Visualizing body-inclusive $Re = 5 \times 10^3$ and $Re = 5 \times 10^4$} 
\label{res:5k_50k_bi}

\begin{figure}
\centering
\includegraphics[width=0.9\linewidth, keepaspectratio]{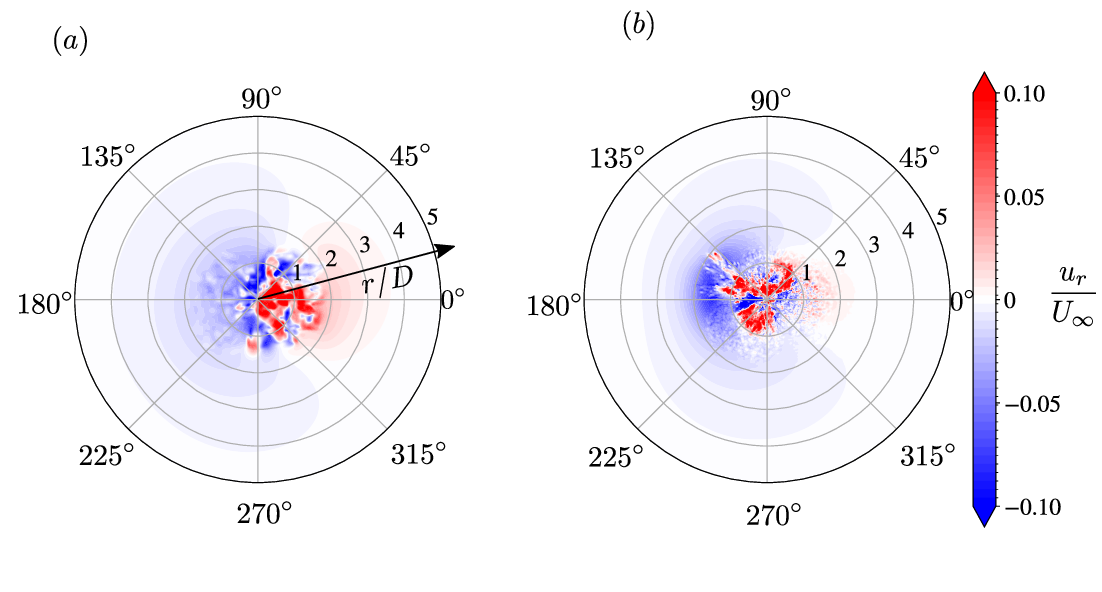}
\caption{Instantaneous snapshot of radial velocity at $x/D=10$ for (a) 5kBI and (b) 50kBI, highlighting differences in the smallest scales.}
\label{fig:5k_50k_bi_snap}
\end{figure}

In this section, we compare SPOD spectra and wake statistics between 5kBI and 50kBI to establish coherent structure similarity prior to high-Re initialization. Figure~\ref{fig:5k_50k_bi_snap} shows contours of instantaneous radial velocity at $x/D = 10$. Both 5kBI and 50kBI exhibit a similar radial extent of the wake. However, 50kBI displays finer structures near the centerline, consistent with the Kolmogorov hypothesis that smaller scales become increasingly prevalent at higher Reynolds numbers due to a larger inertial cascading range.

\subsection{Coherent structures}

\begin{figure}
\centering
\includegraphics[width=\linewidth, keepaspectratio]{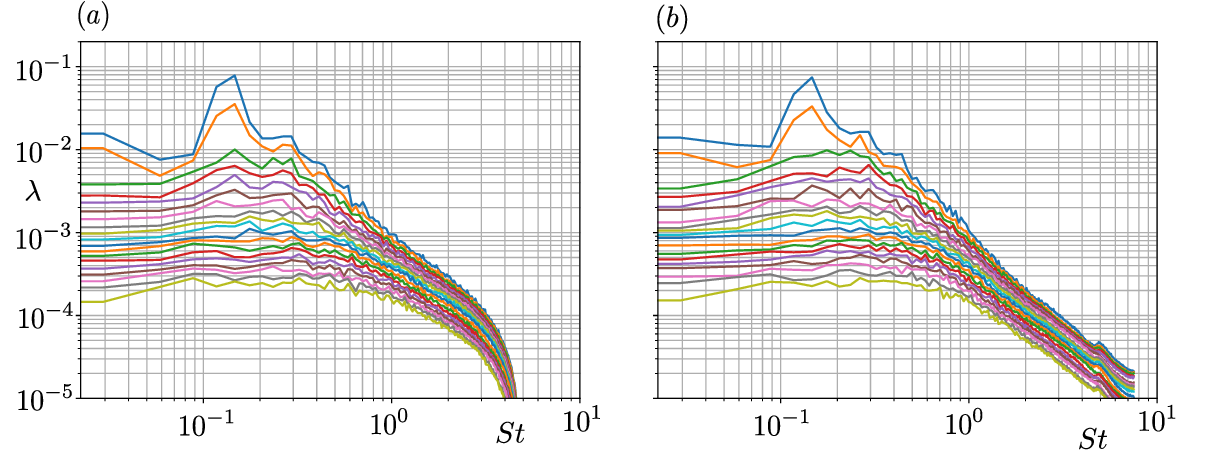}
\caption{SPOD eigenspectra at $x/D = 10$ for (a) 5kBI and (b) 50kBI.}
\label{fig:5k_50k_eigspec}
\end{figure}

\begin{figure}
\centering
\includegraphics[width=\linewidth, keepaspectratio]{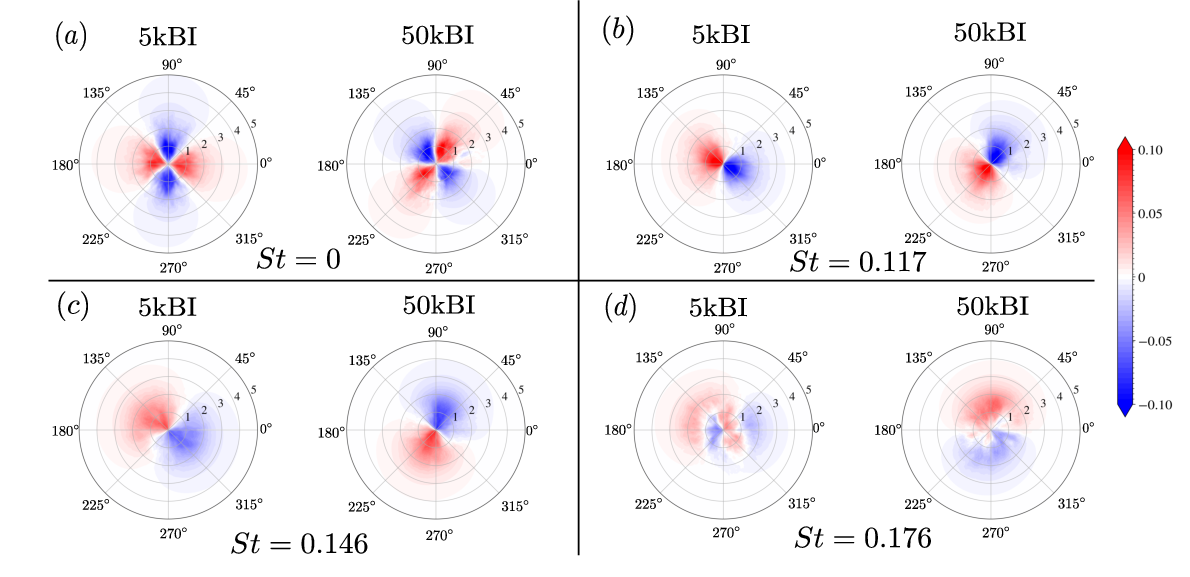}
\caption{Real part of leading SPOD eigenmodes for radial velocity at $x/D = 10$ for 5kBI and 50kBI at low frequencies.}
\label{fig:5k_50k_eigmodes}
\end{figure}

Figure~\ref{fig:5k_50k_eigspec} presents the SPOD eigenspectra for 5kBI and 50kBI. The low-frequency content in both cases is nearly identical, with a prominent peak at the disk vortex shedding frequency, $St = 0.146$. At higher frequencies, 50kBI exhibits slightly greater energy content, consistent with broader spectral support typically observed in higher Reynolds number flows.

The corresponding eigenmodes, shown in Figure~\ref{fig:5k_50k_eigmodes}, exhibit structural similarity at low frequencies. As observed by Nidhan et al.~\cite{nidhan_spectral_2020}, the double helix (DH) mode is dominant at $St = 0$, while the vortex shedding (VS) mode is most prominent at $St = 0.146$. Eigenmodes of the azimuthal and streamwise velocity components (not shown for brevity) also exhibit similar trends between the two cases. Readers are referred to Nidhan et al.~\cite{nidhan_spectral_2020} for a detailed analysis of these modes.

\subsection{Wake evolution}

\begin{figure}
\centering
\includegraphics[width=\linewidth, keepaspectratio]{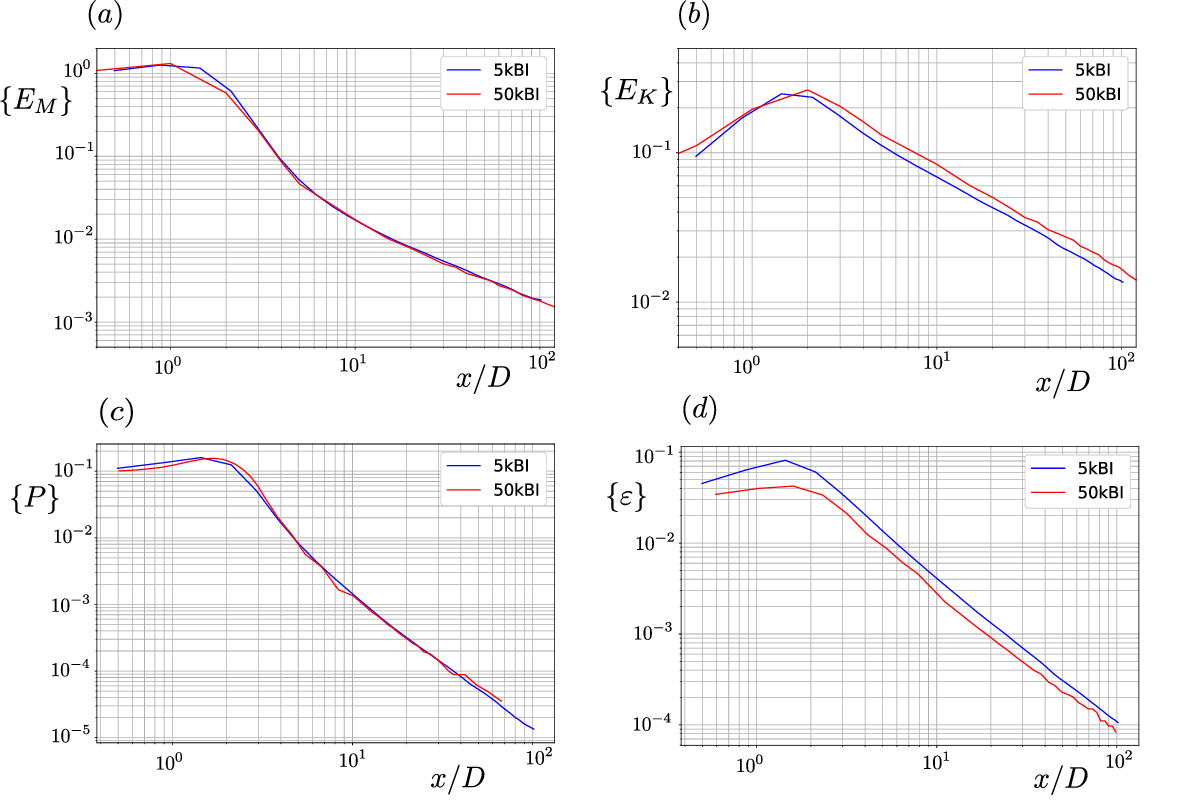}
\caption{Wake statistics as a function of downstream distance for 5kBI and 50kBI: (a) Area-integrated mean kinetic energy (ai-MKE), (b) area-integrated turbulent kinetic energy (ai-TKE), (c) area-integrated production rate $\mathcal{P}$, and (d) area-integrated dissipation rate $\varepsilon$.}
\label{fig:5k_50k_bi_stat}
\end{figure}

Figure~\ref{fig:5k_50k_bi_stat} shows the downstream evolution of area-integrated wake statistics. The area-integrated mean kinetic energy (ai-MKE) and production rate evolve similarly in both cases. However, 50kBI exhibits slightly higher area-integrated turbulent kinetic energy (ai-TKE) and marginally lower area-integrated dissipation. This discrepancy is likely due to the enhanced viscous effects at lower Reynolds numbers, leading to greater energy dissipation and reduced turbulent kinetic energy in the 5kBI case.

These observations suggest that when initializing a high-$Re$ simulation using low-$Re$ flow fields or their reconstructions, the downstream adjustment of ai-TKE will depend on the amount of retained turbulent kinetic energy and Reynolds stresses. The production term is particularly critical, as it governs energy transfer from the mean to turbulent scales.

\subsection{Reconstructed data from $Re = 5 \times 10^3$}

\begin{figure}
\centering
\includegraphics[width=0.85\linewidth, keepaspectratio]{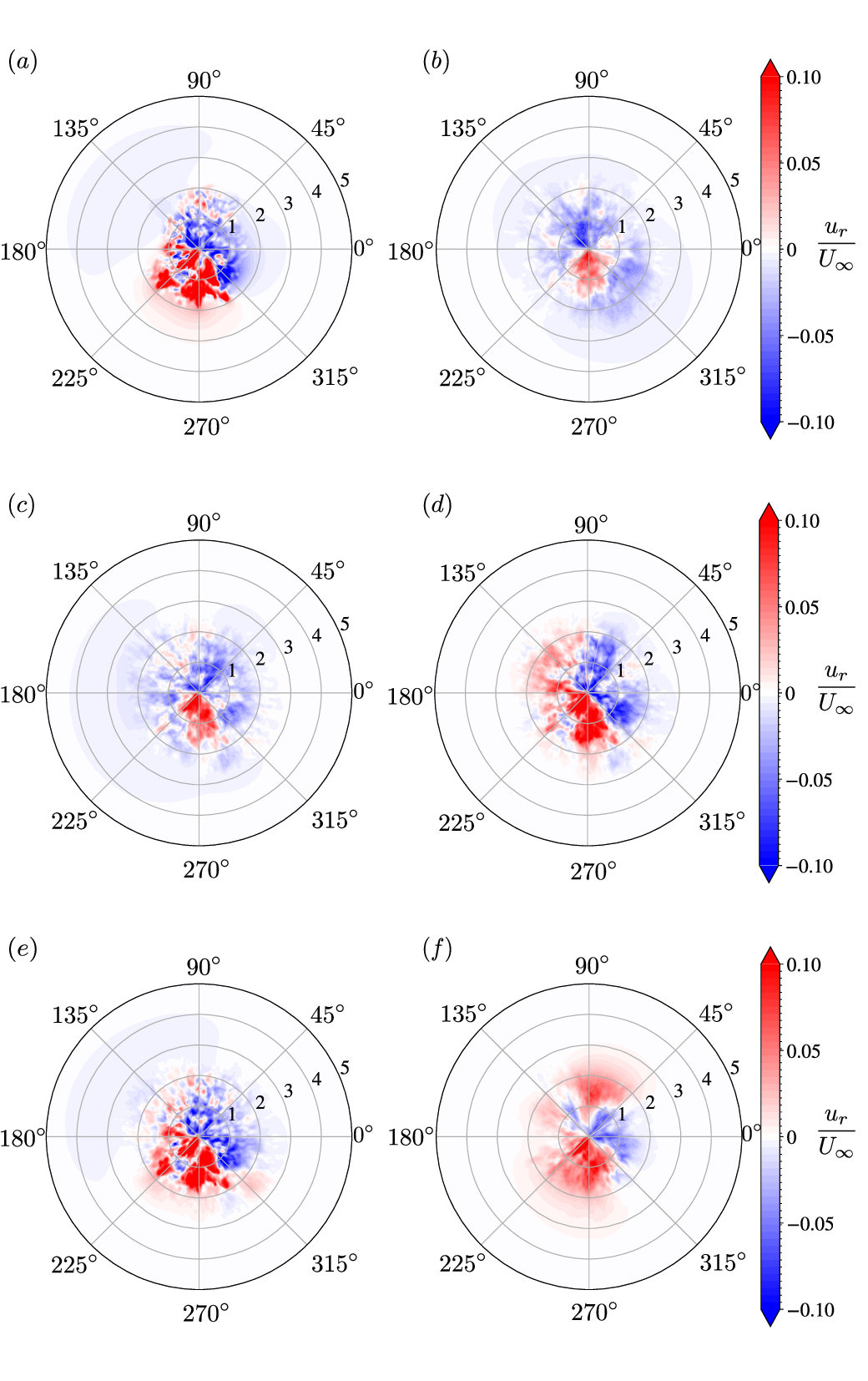}
\caption{Instantaneous contours of radial velocity for: (a) 5kBI/SRH-NAll, (b) SRH-N1, (c) SRH-N2, (d) SRH-N3, (e) SRH-N6, (f) SRH-St0103.}
\label{fig:srh_ur_comp}
\end{figure}

\begin{figure}
\centering
\includegraphics[width=\linewidth, keepaspectratio]{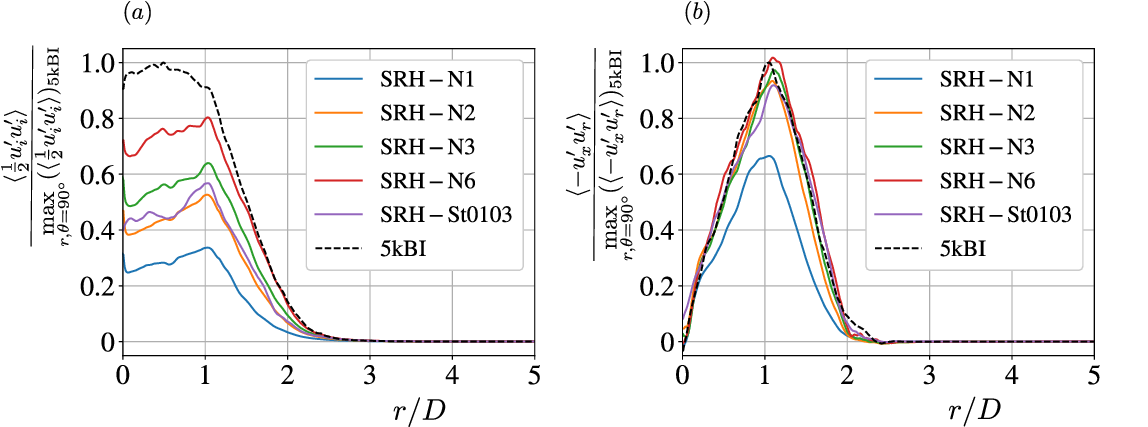}
\caption{(a) ai-TKE of reconstructed cases normalized by maximum 5kBI value. (b) Streamwise-radial Reynolds stress $\langle -u_x'u_r' \rangle$ normalized by 5kBI maximum, plotted along $\theta = 90^\circ$.}
\label{fig:srh_rey_tke_comp}
\end{figure}

Figure~\ref{fig:srh_ur_comp} shows the radial velocity contours for the reconstructed SRH cases (b–f) compared to the original 5kBI field (a). As more SPOD modes are included in the reconstruction, the resemblance to 5kBI improves. Notably, the SRH-St0103 case (f), which includes all modes but is restricted to $0.1 < St < 0.3$, can be interpreted as a bandpass-filtered version of the original field.

Figure~\ref{fig:srh_rey_tke_comp}(a) compares the ai-TKE content along $\theta = 0^\circ$. Low-rank reconstructions such as SRH-N1, SRH-N2, and SRH-N3 retain less than half the TKE, similar to the band-limited SRH-St0103 case. However, as seen in Figure~\ref{fig:srh_rey_tke_comp}(b), the streamwise-radial Reynolds stress is much better preserved across cases. 

It is also notable that Reynolds stress differences between SRH-N1 and SRH-N2 are substantial, while SRH-N2, SRH-N3, and SRH-N6 show close agreement. This suggests that as few as two leading SPOD modes are sufficient to approximate the Reynolds stresses with reasonable accuracy. Since production depends on Reynolds stress and mean velocity gradients (which are identical across cases), good retention of Reynolds stress implies improved prediction of downstream evolution of mean defect velocity, even when the initial TKE is under-predicted.

These results demonstrate that the dominant coherent structures and their associated energy pathways are robust across an order-of-magnitude change in Reynolds number. This provides a compelling justification for initializing high-Reynolds number simulations of the present bluff-body wake  using low $\Rey$ flow fields or their low-rank reconstructions, provided that Reynolds stress structures are retained.

\section{SRH body-exclusive simulations at $Re = 5 \times 10^4$}
\label{res:50kbe}

\subsection{Wake evolution}




\begin{figure}
\centering
\includegraphics[width=\linewidth, keepaspectratio]{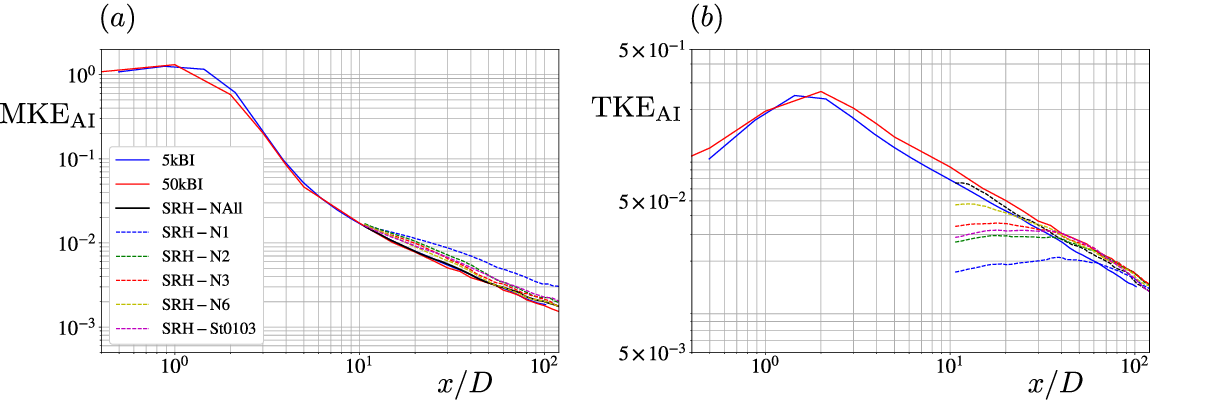}
\caption{Downstream evolution of (a) area-integrated mean kinetic energy (ai-MKE) and (b) area-integrated turbulent kinetic energy (ai-TKE) for all SRH cases and 50kBI.}
\label{fig:ai_mke_tke_all}
\end{figure}

Figure~\ref{fig:ai_mke_tke_all} shows the downstream variation of ai-MKE and ai-TKE for all SRH cases in comparison to 50kBI. While the SRH-N1 case deviates from the expected ai-MKE decay, all other SRH cases closely follow the trend observed in 50kBI. In terms of ai-TKE, all SRH simulations eventually converge to the same decay rate as 50kBI. The time or distance required to achieve this convergence is inversely related to the number of modes included in the initial reconstruction. Notably, SRH-N2 and higher-rank reconstructions, as well as the band-limited SRH-f13 case, successfully recover both ai-MKE and ai-TKE decay characteristics of the full 50kBI simulation after an initial adjustment period.

These trends are driven by the behavior of the Reynolds stress terms, which regulate energy transfer via production from the mean flow to turbulence. Since most SRH cases (except SRH-N1) retain the bulk of the Reynolds stress structure (see Figure~\ref{fig:srh_rey_tke_comp}b), the production terms are similarly well preserved. For SRH-N1, however, reduced Reynolds stresses result in lower production, which slows the energy transfer and leads to a shallower MKE decay explaining the deviation in Figure~\ref{fig:ai_mke_tke_all}(a).

In contrast, the ai-TKE evolution highlights how an initial under-prediction of TKE in low-rank reconstructions leads to an ``adjustment period," during which production drives up the TKE until it reaches the expected decay trend. After this transition, the SRH simulations behave consistently with fully resolved $Re = 5 \times 10^4$ flows, as further validated using SPOD in the following section.

\subsection{Validation using SPOD}

\begin{figure}
\centering
\includegraphics[width=\linewidth, keepaspectratio]{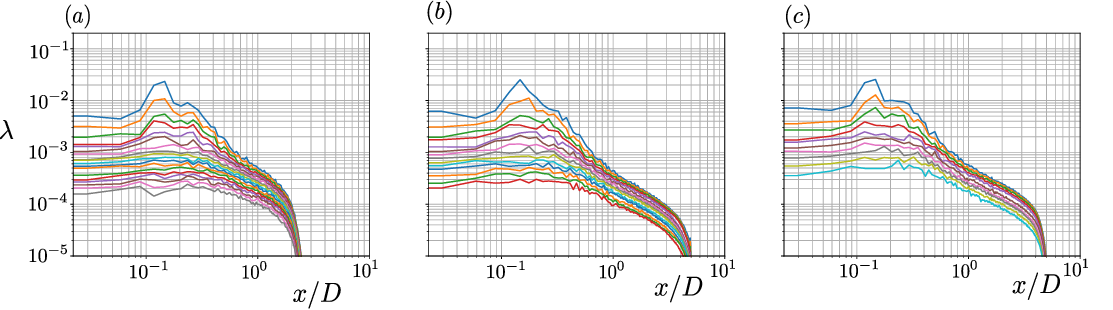}
\caption{SPOD eigenspectra of radial velocity at $x/D=30$ for (a) 5kBI, (b) 50kBI, and (c) SRH-NAll.}
\label{fig:eigsp_30_all}
\end{figure}

\begin{figure}
\centering
\includegraphics[width=0.8\linewidth, keepaspectratio]{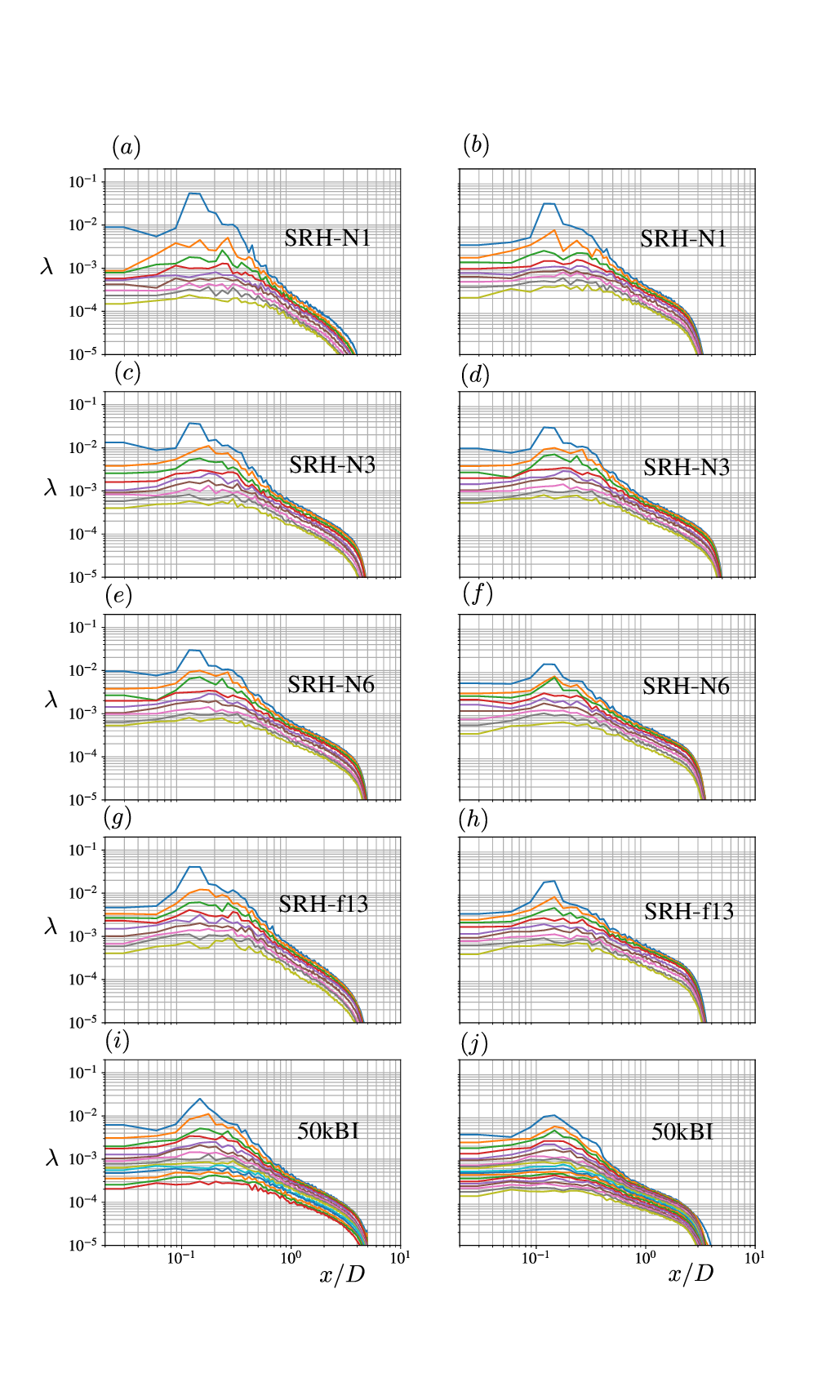}
\caption{SPOD eigenspectra of radial velocity at $x/D=30$ (left column) and $x/D=60$ (right column) for SRH cases (not all shown for brevity).}
\label{fig:eigsp_srh_all}
\end{figure}

To assess the fidelity of the SRH simulations, we first compare the SPOD eigenspectra at $x/D = 30$ for 5kBI, 50kBI, and SRH-NAll cases. As illustrated in Figure~\ref{fig:eigsp_30_all}, the eigenspectra for these three cases exhibit strong agreement, particularly in the low-frequency range where coherent structures dominate. Notably, SRH-NAll shows similarity with 5kBI in the dominant low-frequency modes and with 50kBI in the higher-frequency content. This dual agreement suggests that the SRH methodology effectively bridges the spectral characteristics of the low-resolution inflow and the high-Reynolds-number target flow. The preservation of large-scale energetic modes and eventual development of higher-frequency content indicates that SRH can accurately reproduce the essential flow physics downstream of the adjustment region.


Figure~\ref{fig:eigsp_srh_all} further compares the SPOD eigenspectra of the radial velocity at $x/D = 30$ and $x/D = 60$ for four SRH configurations against the 50kBI benchmark. In all SRH cases, the spectra exhibit extended high-frequency tails comparable to those in 50kBI, indicating that small-scale turbulent structures are present and persist downstream. Moreover, the prominent low-frequency peaks originating from the 5kBI inflow are retained, demonstrating the robustness of the SPOD-based inflow construction in seeding the correct large-scale dynamics. Variations in spectral amplitude across cases can be attributed to two main factors: (i) the limited number of inflow modes retained in the SRH reconstructions, which concentrates more energy in the leading modes, and (ii) the reduced number of snapshots used in the SPOD analysis, resulting in fewer temporal averaging blocks ($n_\mathrm{blk} = 9$ in SRH vs. 19 in 50kBI). Nonetheless, increasing the number of retained inflow modes leads to improved spectral alignment with the 50kBI reference, highlighting the importance of inflow content in capturing the full spectral and coherent structure of the turbulent wake.

\subsection{Computational cost analysis}

One of the primary motivations for the SRH approach is the significant reduction in computational expense. The 5kBI simulation required approximately 22,000 compute hours per flow-through on 128 cores. With three flow-throughs used for data collection (following an initial transient of one flow-through), the total compute time for 5kBI was about 88,000 hours.

Each SRH simulation required only 3,200 compute hours per flow-through on 128 cores. Simulations were run for approximately 2.5 flow-throughs (1 for reaching statistical stationarity, 1.5 for data collection), resulting in a total cost of roughly 10,000 compute hours per SRH case.

The surprisingly low compute time for SRH simulations is attributed to a coarser time step enabled by the computational domain starting at $x/D = 10$, away from the high-gradient near-wake region. The total cost of a hybrid simulation (5kBI + one SRH run) is thus approximately 98,000 compute hours on 128 cores for 2.5 flow-throughs of high-Re data.

In comparison, the full 50kBI simulation by \citet{chongsiripinyo_decay_2020} required approximately 600,000 compute hours on 512 cores for a similar time span. This corresponds to more than an 80\% reduction in total compute cost using the SRH approach without sacrificing spatial resolution or fidelity in the far wake.

\section{Discussion}
\label{res:discuss}

Hybrid simulations offer a powerful approach to reduce computational costs by decomposing a full-body-inclusive simulation into a short body-inclusive precursor simulation followed by an extended body-exclusive continuation simulation. This strategy has previously been used to handle computationally expensive turbulent flows \citep{vandine_hybrid_2018, ortiz2021high}. The present study further enhances the hybrid simulation methodology by incorporating Spectral Proper Orthogonal Decomposition (SPOD) as a filtering and reconstruction tool to prescribe inlet boundary conditions and reduce the number of required modes for flow reconstruction.

A key outcome of this study is the demonstration of scale similarity in the dominant coherent structures across Reynolds numbers. Specifically, the low-frequency energetic modes of the wake behind a disk show minimal variation between $Re = 5 \times 10^3$ and $Re = 5 \times 10^4$. This Reynolds number independence of the dominant SPOD modes enables the successful initialization of high-Reynolds number body exclusive simulations using filtered data from low $\Rey$ body inclusive simulations. The resulting SRH simulations achieve a reduction in computational cost of over 80\% compared to the full 50kBI simulation, while preserving spatial resolution and key flow dynamics.

The effectiveness of this hybrid strategy is validated through the wake evolution of SRH cases. The adjustment distance defined here as the streamwise region required for the SRH simulations to develop small-scale features and match the $Re = 5 \times 10^4$ wake characteristics, was found to vary inversely with the number of SPOD modes used for reconstruction. Cases using more modes (e.g., SRH-N3 and SRH-N6) showed faster convergence in mean and turbulent kinetic energy profiles to the baseline 50kBI results. The area-integrated MKE and TKE plots indicate that all cases except SRH-N1 eventually converge to the correct decay trends, confirming the sufficiency of large-scale content in the SPOD-reconstructed inlet data.

This behavior is further explained by analyzing Reynolds stress distributions. All SRH cases except SRH-N1 adequately capture the key Reynolds stress components, ensuring that mean-to-turbulence energy transfer occurs correctly. The underestimation of these stresses in SRH-N1 leads to a delay in the decay of mean kinetic energy, underscoring the importance of capturing the appropriate production mechanisms. Once the small scales have developed and the energy budget is balanced, the flow transitions into a statistically stationary regime representative of the higher Reynolds number, as verified by SPOD eigenspectra matching those of the 50kBI simulation.

The results presented here reinforce the idea that a significant amount of turbulent kinetic energy and Reynolds shear stress are captured by a relatively small number of large-scale modes, aligning with recent observations in turbulent flow analysis \citep{nidhan_spectral_2020}. This offers a strong justification for using SPOD-based reconstructions not only in numerical simulations but also for coupling with experimental datasets, which often lack full spatial and temporal resolution. Future work could extend this methodology to high Reynolds number stratified flows, where simulating far-wake decay remains a significant challenge in traditional body-inclusive simulations. The SRH method proposed here leverages the  low-rank behavior of a bluff-body wake.  Evaluation of  its applicability to other wake generators, e.g., slender bodies at different values of angle of attack,  is  worth further research. Additionally, the SRH approach may facilitate high-fidelity simulations of complex geometries at high Reynolds numbers by significantly reducing computational cost and enabling targeted flow reconstruction. 

\section{Acknowledgment}

We are pleased to acknowledge the funding provided by ONR grant N000142012253.


\bibliographystyle{elsarticle-harv} 
\bibliography{hybrid_spod_v5}

\end{document}